\documentclass[iop]{emulateapj}
\usepackage{natbib}
\usepackage{amsmath}
\usepackage{graphicx}
\usepackage{dcolumn}
\usepackage{bm}

\shorttitle{Formation of current singularities}
\shortauthors{I.~J.~D.~Craig \& F.~Effenberger}

\begin{document}

\title{Current Singularities at Quasi-Separatrix Layers and Three-Dimensional Magnetic Nulls}

\author{I.~J.~D.~Craig\altaffilmark{1} and Frederic Effenberger\altaffilmark{1}}

\altaffiltext{1}{Department of Mathematics, University of Waikato,
  P.B. 3105, Hamilton, New Zealand}

\begin{abstract}
The open problem of how singular current structures form in line-tied,
three-dimensional magnetic fields is addressed. A Lagrangian
magneto-frictional relaxation method is employed to model the field
evolution towards the final near-singular state. Our starting point is
an exact force-free solution of the governing magnetohydrodynamic
(MHD) equations which is sufficiently general to allow for topological
features like magnetic nulls to be inside or outside the computational
domain, depending on a simple set of parameters. Quasi-separatrix
layers (QSLs) are present in these structures and together with the
magnetic nulls, they significantly influence the accumulation of
current. It is shown that perturbations affecting the lateral
boundaries of the configuration lead not only to collapse around the
magnetic null, but also to significant QSL currents. Our results show
that once a magnetic null is present, the developing currents are
always attracted to that specific location and show a much stronger
scaling with resolution than the currents which form along the QSL. In
particular, the null point scalings can be consistent with models of
``fast'' reconnection. The QSL currents also appear to be unbounded but
give rise to weaker singularities, independent of the perturbation
amplitude.
\end{abstract}



\section{Introduction}
\label{sec:intro}
Magnetic reconnection is the mechanism that allows topological change
in weakly resistive magnetic plasmas such as the solar corona. For
reconnection to be effective, huge currents involving steep field
gradients must be present. How these near-singular current structures
develop is not fully understood, but it is generally recognised that
topological features of the field -- null points and separators --
should play a decisive role \citep{Lau-Finn-1990, Priest-Titov-1996,
  Pontin-Craig-2006}. More physically, these features are likely to
provide sites for the quiescent heating of the corona and the rapid
energy release in solar flares \citep{Parker-1972}.

One route to understanding reconnection is to examine the
eigenstructure of 3D magnetic nulls. This provides a field skeleton
that comprises ``spine'' lines and ``fan'' planes which accumulate
current when the null is suitably perturbed. For instance, bending the
spine of an isolated X-point, leads to a current layer aligned to the
fan i.e. ``fan'' reconnection \citep[for a review on topological
  aspects in reconnection see, e.g.,][]{Longcope-2005}. However, a
different current structure emerges when the fan plane is distorted:
this leads to ``spine'' reconnection in which quasi-cylindrical tubes
of current become localized to the spine axis
\citep{Craig-Fabling-1996}.

Current sheet formation and reconnection can also take place in the
absence of a null. The key feature in this case is the
``quasi-separatrix layer'' or QSL for short
\citep{Priest-Demoulin-1995,Titov-etal-2002,Demoulin-2006,Aulanier-etal-2006}.
This is a region of rapid variation in field-line connectivity which
can be thought of as being of geometrical \citep{Titov-Hornig-2002}
rather than topological significance.  The simplest example is
provided by a line-tied planar X-point threaded by uniform axial
field. The QSL extends between the upper and lower planes $ z = \pm L
$ (say) and replaces the separatrix surfaces of the purely planar
X-point. All points on the QSL are connected Alfv\'enically, but there
is no unique point, like a magnetic null, on which currents can
accumulate \citep{Galsgaard-2000,Craig-Pontin-2014}.

There is observational evidence that current formation and
reconnection involving coronal active region outflows are connected to
QSLs \citep{Baker-etal-2009, Guo-etal-2013}. Models of Solar flares
\citep{Demoulin-etal-1996, Zhao-etal-2014} and coronal mass ejections
also suggest that QSLs can play a decisive role in the initiation of
such eruptions \citep{Schrijver-etal-2011,Savcheva-etal-2012}. More
theoretically, since QSLs provide strong layers of currents, they
must be regarded as prime sites for particle acceleration in the
active corona \citep[cf.][]{Heerikhuisen-etal-2002, Stanier-etal-2012}.

How QSLs influence reconnection is not entirely clear. Opinion is
divided on whether disturbances which include shifts of the line-tied
boundary can initiate a collapse to an ideal current singularity -- as
they do in null point reconnection -- or whether very steep but finite
current distributions are obtained \citep{Titov-etal-2003,
  Galsgaard-etal-2003}. What is known, at least for magnetic X-points,
is that by strengthening the effects of axial line-tying, either by
increasing the axial field or shortening the distance between the
upper and lower line-tied boundaries, current localization can be
inhibited \citep{Craig-Pontin-2014}. This leads to the strongest
current accumulation around the outer boundaries of the X-point, which
is unfavourable to rapid reconnection and artificial in terms of a
computational reconnection experiment.

In order to prescribe an initial field in which QSL structures or
magnetic nulls are present, a potential field generated by submerged
magnetic monopoles is often considered. Studies of these magnetic
structures, driven by slow lower-boundary photospheric motions, show
strong currents developing along the QSL with no signs of saturation
in the accessible numerical resolution regime
\citep{Aulanier-etal-2005,Effenberger-etal-2011}. The strong gradients
due to the monopoles which are situated closely below the lower
boundary can, however, present numerical difficulties that may result,
artificially, in the largest currents forming close to that boundary.

The aim of the present study is to compare QSL current structures with
those in isolated magnetic nulls. In contrast to previous studies
based on the submerged monopole approach, we use line-tied
analytically prescribed fields that allow for a continuous
transformation between null-point and QSL equilibria. This allows null
point and QSL currents to be treated on the same physical footing.
In \S\ref{sec:fields} we discuss the form of the general force-free
equilibrium field we employ as the basis of our study and detail how
perturbations can lead to current formation. In \S\ref{sec:comp} we
will present results from relaxation runs performed with both a
linearized, potential version of this field, and the more general
force-free field. We finally discuss our findings in \S\ref{sec:discussion}.

\section{Generalized QSL and X-point fields}
\label{sec:fields}
We consider an equilibrium field $\mathbf{B}(\mathbf{r})$ defined
within the interior of the rectangular domain $\cal{D}$ and line-tied
on all the bounding surfaces. This field is subject to some finite
amplitude disturbance that displaces the footpoints of the
equilibrium field altering the magnetic topology. Magnetic
reconnection allows the perturbed field to ``relax'' dynamically into
a new equilibrium which is topologically different from the initial
disturbed field. However, if resistive effects are absent but some
other form of damping is present, the perturbed field has to relax
without topological change. In this case, a near-singular
configuration can emerge, comprising strong, highly localized current
densities.  It is the properties and definition of these near-singular
``relaxed'' fields that are the focus of the present study.

Our starting point is an initial field $\mathbf{B}(\mathbf{r})$ with
the components
\begin{align}
B_x &= \kappa\mu x \cos(\mu z - b) + (1-\kappa)\mu y \sin(\mu z - b)
\\ B_y &= (1-\kappa)\mu y \cos(\mu z - b) - \kappa\mu x \sin(\mu z -
b) \\ B_z &= -  \sin(\mu z - b) \,.
\label{eq:B}
\end{align}
We assume that field intensities and distances are scaled according
to typical coronal values, for example $10^2$~G for the magnetic field
and $ 10^{9.5}$~cm for the coronal size scale. The domain ${\cal D}$
is then the region $-1,\,\le x,y,z\le\,1$.

The initial field is defined by the three parameters, $ \mu, \kappa$
and $b$ and has the following properties:
\begin{align}
\nabla\cdot\mathbf{B}  &= 0 \\
\nabla\times\mathbf{B} &= \mu  \mathbf{B} \\
\nabla^2\mathbf{B}     &= -\mu^2\mathbf{B} \,.
\label{eq:Bcond}
\end{align}
We see that $\mu$ accounts for a rotation phase, while $\kappa \ne 1/2$
allows for rotational asymetries about the $z$-axis in the field. The
null is located at the point
\begin{align}
\mathbf{r}_p = \frac{b}{\mu}\,\hat{\mathbf{z}}
\label{eq:null-pos} 
\end{align}
and, for fixed $\mu $, can be shifted outside the field domain by  
adjusting $b$. The field, being force-free, is considerably more
general than potential fields due to the presence of parallel
currents.

\subsection{Related potential fields}   
\label{sec:pot-fields}
Potential fields can be extracted from the general field (\ref{eq:B})
by formally regarding $\mu$ and $b$ as sufficiently small parameters.
We then obtain the linear potential field
\begin{align}
\mathbf{P}_{1} = (\kappa\mu x,\,(1-\kappa)\mu y,\,b -\mu z)\,. 
\label{eq:Bsimp0}
\end{align}
We can avoid redundant parametrization by taking $\kappa=1/\mu$:
\begin{align}
\mathbf{P}_{2} = (x,\,(\mu -1)y,\,-\mu z + b)\,.
\label{eq:Bsimp}
\end{align}
In this case we should regard $\mu$ as a proxy for the isotropy
parameter. Note that, although the potential forms allow considerable
simplification, they can be expected to provide a reasonable guide to
the current accumulation properties of the more general field
(\ref{eq:B}) at least in regions close to the null. This point is
revisited in \S\ref{sec:comp} below.

\subsection{Spines, Fans and QSLs} 
\label{sec:spine-fan-qsl}
In our analysis, we consider only perturbing fields $\mathbf{B}_{p}$
that disturb the lateral boundaries of the domain. To illustrate the
effect of these disturbances, consider the simple form
\begin{align}
\mathbf{B}_{p} &= (0,\,a_1 x,\,a_2 x)  
\label{eq:Apert}
\end{align}
in the superposition
\begin{align}
\mathbf{B}_s &= \mathbf{P}_2  + \mathbf{B}_p \nonumber\\
             &= (x,\,(\mu - 1)y+a_1x,\,b - \mu z + a_2x)\,.  
\label{eq:asup}
\end{align}
Provided that $a_1$ or $a_2$ are non-zero, the perturbation is finite
on the boundary points $x = \pm 1$. The field line equations
$\mathbf{dr}\propto \mathbf{B}_s$ in the case of a null point field
give (assuming $ \mu \ne 0 $):
\begin{align}
x^{1 - \mu} \left(y - \frac {a_1 x} {2 - \mu}\right) = C_1,   \quad  x^{\mu} \left(z -
\frac {b} {\mu}  - \frac {a_2 x} {\mu +1}\right) = C_2 \,, 
\label{eq:fieldeq}
\end{align}
where $C_1$ and $C_2$ are constants. The separatrices are defined by
field lines that thread the null, i.e. by setting $C_1=C_2=0$. We
obtain the ``fan'' plane $x = 0$ and the ``spine'' line
\begin{align}
y =  \frac {a_1 x} {2 - \mu},  \quad z = \frac {b} {\mu} + \frac {a_2 x} {\mu +1} \,. 
\label{eq:sep}
\end{align}
We see that the spine line of the unperturbed field -- the $x$-axis
in the case $b = 0$ -- becomes tilted due to the perturbation. This
generates currents of magnitude $\sqrt{a_1 ^2 + a_2 ^2}$ within the
fan. When no resistivity is present, these initial currents localize --
and eventually blow up -- in the vicinity of the null point.

The fan-spine structure breaks down when the null point is absent. In
the simplest case of a planar field with no axial component
($\partial_z = \mu = b = 0 $) the separatrices (obtained by the first
of Eqs.~\ref{eq:fieldeq}) are just the two planes $x = 0$ and $y = 0$,
the latter being tilted through the angle $ \tan \theta = a_1 / (2 -
\mu )$ \citep[see, e.g.,][]{Craig-Pontin-2014}. The null point is now
extended to a null line defined by the intersection of the two planes.

In the case of a constant finite axial field ($b \ne 0$, $\mu=0$) the
null line is removed and separatrices cannot be defined. Even so, each
$z$-plane still retains a projected copy of the tilted separatrix
planes of the perturbed planar X-point. Reconnection again requires
currents localized towards the $z$-axis but now involves field lines
whose ends are anchored across different $z$-planes. Line-tying the
axial field on $z = \pm1$, however, breaks the symmetry $\partial_z =
0$ and makes the geometry fully three-dimensional. This is the
topology of the QSL: no null is present but magnetic stresses can
accumulate due to steep field gradients centred on the $z$-axis.

Finally, we turn to the field line equations in the case $ \mu = 0 $,
that is, the simplest case of a QSL field (i.e. $B_z=b$), without a null:      
\begin{align}
x \left(y - \frac {a_1 x} {2}\right) = C_3, \quad  
x \exp\left(\frac{a_2 x \, - z}{b}\right) = C_4  \,.
\label{eq:fieldeqnull}
\end{align}
We see that there is strong exponential dependency on the $C_4$ field
line even when the perturbation amplitude vanishes ($ a_2 = 0 $).
This dependency reflects the geometrical squashing factors associated
with equilibrium field lines when no null is present 
\citep{Titov-2007}.    

\subsection {Squashing factors for QSLs}
As a complement to the computational study of \S\ref{sec:comp}, it is
instructive to determine QSL squashing factors for the case where
$a_2$ vanishes but $a_1$ remains finite. In this case we regard the
upper and lower boundaries as adjustable planes $z = \pm z_m$ (say)
and use the field line equations (\ref{eq:fieldeqnull}) to relate the
upper and lower footpoint locations. Denoting the footpoint
coordinates at the upper and lower boundaries with $x_\pm$ and $y_\pm$
we have that
\begin{equation}
\left ( \begin{array} {c} x_{+} \\ y_{+} \end{array} \right)  =
\left ( \begin{matrix}  \exp (2  \, \frac  {z_m} {b}) & 0 \vspace{0.1cm}\\ 
a_1   \sinh (2 \frac {z_m} {b} )  &  \, \, \exp (- 2 \frac {z_m}
{b})   \end{matrix} \right)  
\left ( \begin{array} {c} x_{-} \\ y_{-} \end{array} \right) .
\end{equation}
These expressions can be used to determine the invariant squashing
factor $Q$ along the tube axis as defined in \citep{Titov-2007}. We
find that
\begin{equation}
Q = 2 \, \cosh (4 \frac {z_m} {b} ) + a_1^2 {\sinh ^2}  (2 \frac {z_ m} {b} )  
\end{equation}
For modest perturbation amplitudes (i.e.\ $|a_1| < 1$) the squashing
factor is determined mainly by the form of the equilbrium field:
specifically $Q$ increases rapidly with the tube length ($2 z_m$) but
decreases with the strength of the axial field. Given these properties
it seems natural to suppose that current accumulation in the QSL might
reflect this behaviour. There is some numerical evidence that longer
tubes can lead to stronger current localizations along the tube axis
\citep{Craig-Pontin-2014} but, to our knowledge, a direct link between
QSL currents and squashing factors has not yet been established.

\section{Computational results for current formation}
\label{sec:comp}
The previous discussion suggests that reconnection can be expected
within QSL configurations in response to foot-point displacements of
the lateral boundaries, like those given by
Equation~(\ref{eq:Apert}). What is less certain is the strength and
location of the reconnection currents within the QSL. One possibility
is that, in addition to the form and amplitude of the foot-point
displacements, current strengths may be moderated by geometric
squashing factors \citep{Titov-2007} associated with the equilibrium
configuration.  More specifically, given that the present
computational setup reflects the disturbed field ${\bf B}_s$ of
Section~\ref{sec:spine-fan-qsl}, we expect to see steep field
gradients aligned to the $z$-axis of the domain.

To investigate further, we now compare the current structures of
perturbed magnetic field equilibria, both in the presence and in the
absence of a null point. We use the magneto-frictional Lagrangian
scheme of \citet{Craig-Pontin-2014} which models line-tying by fixing
fluid elements on the boundary of the domain ${\cal D}$. In practice
we adopt a uniform distribution of $N^3 $ fluid particles in the
interior of ${\cal D}$ and follow their evolution according to the
Lorentz forces on the plasma. The code is implicit and unconditionally
stable.  The solenoidal condition $\nabla \cdot {\bf B} = 0 $ is
satisfied to machine accuracy and flux and magnetic helicity are
conserved.  Gas pressure forces are neglected in the runs that follow.

The perturbation field we adopt is based on the simple model
(cf. Eq.~\ref{eq:Apert}):
\begin{align}
\mathbf{B}_{p} = [& A \sin(\pi x/2) \cdot
  (1-y^2)\cdot(1-z^2)\nonumber\\
&\cdot\exp(-4x^2 - 3 y^2 )] \, \hat {\bf y}
\label{eq:Cpert}
\end{align}
where $A$ denotes the perturbation amplitude. The perturbation
vanishes on all boundary points except $x = \pm 1$. Note that because
the frictional relaxation can lead to divergent current structures the
computed values of certain local variables may be sensitive to
numerical resolution.  This makes it possible to obtain scaling laws
in which the maximum current density in the domain ${\cal D}$ can be
systematically quantified as a function of the resolution, i.e. the
linear number $N$ of fluid particles in the domain.
 
\subsection{Relaxation of the current density}
\label{sec:relax}
\begin{figure}
\noindent\includegraphics[width=0.49\textwidth]{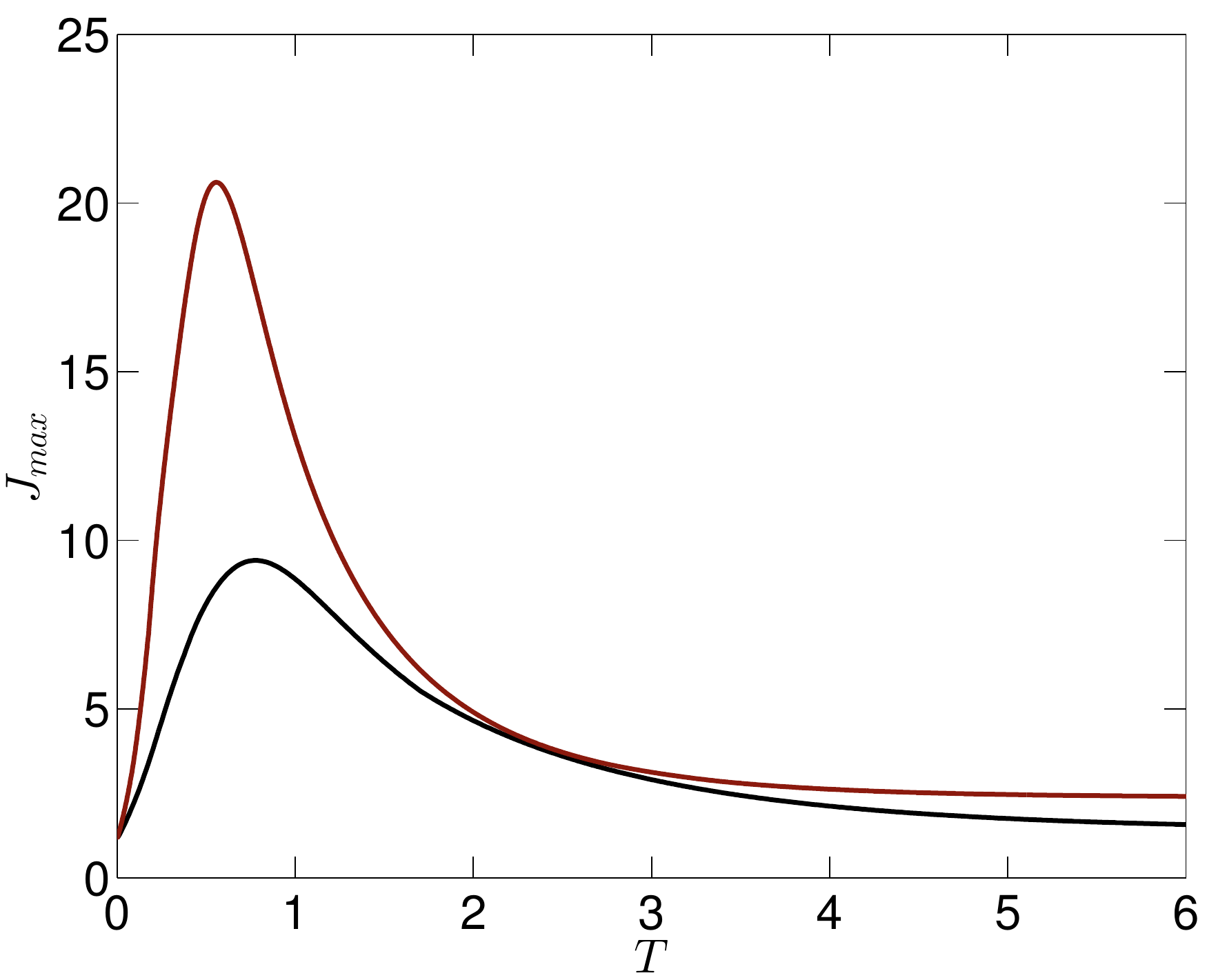}
\caption{Pseudo-time evolution of the maximum current density during
  the relaxation to a force-free equilibrium for two different
  resolutions ($N=61$, black and $N=81$, red). Time $T$ has been
  independently normalized in both cases to give comparable values of
  order one.}
\label{fig:psedotime}
\end{figure}
The relaxation of the maximum current density in the domain for
typical runs based on field $\mathbf{P}_2$ with $b=0.3$ and $\mu=-0.4$
are displayed in Figure~\ref{fig:psedotime} for two numerical
resolutions, namely $N=61$ and $N=81$. The pseudo time parameter $T$
is based on uniform time increments and reflects the number of
iterations in the relaxation. In practice the perturbation amplitude
is chosen so that the initial forces and currents are of order unity
($A = 0.3$ in the present runs) and the computation is halted when
forces are reduced by four orders of magnitude. This protocol allows
well defined scaling laws of the form $J_{max} = a_0 N ^ \alpha$ where
$a_0$ and $\alpha$ are constants and $J_{max}$ is the final maximum
current density in the domain. Detailed scaling laws are discussed in
Section~\ref{sec:scaling} below. 

Returning to Figure~\ref{fig:psedotime}, it is clear that higher
resolution runs are associated with higher relaxed current
densities. Furthermore, an even higher transitory current peak is
visible in both cases. We do not investigate this phenomenon further
in this study, since we are interested mainly in the near-singular,
relaxed state.  It seems worth remarking, however, that {\it resistive}
relaxation in a full MHD configuration often involves inertial
overshoots that lead to oscillatory null-point reconnection
\citep[see, e.g.,][]{Craig-McClymont-1991,Craig-Watson-1992}. It is
likely therefore that the large transient currents in the pseudo time
evolution may reflect the strong initial implosion of the disturbance
field towards the null.

\subsection{QSL versus null-point structure}
\label{sec:struc}
\begin{figure}
\includegraphics[width=0.49\textwidth]{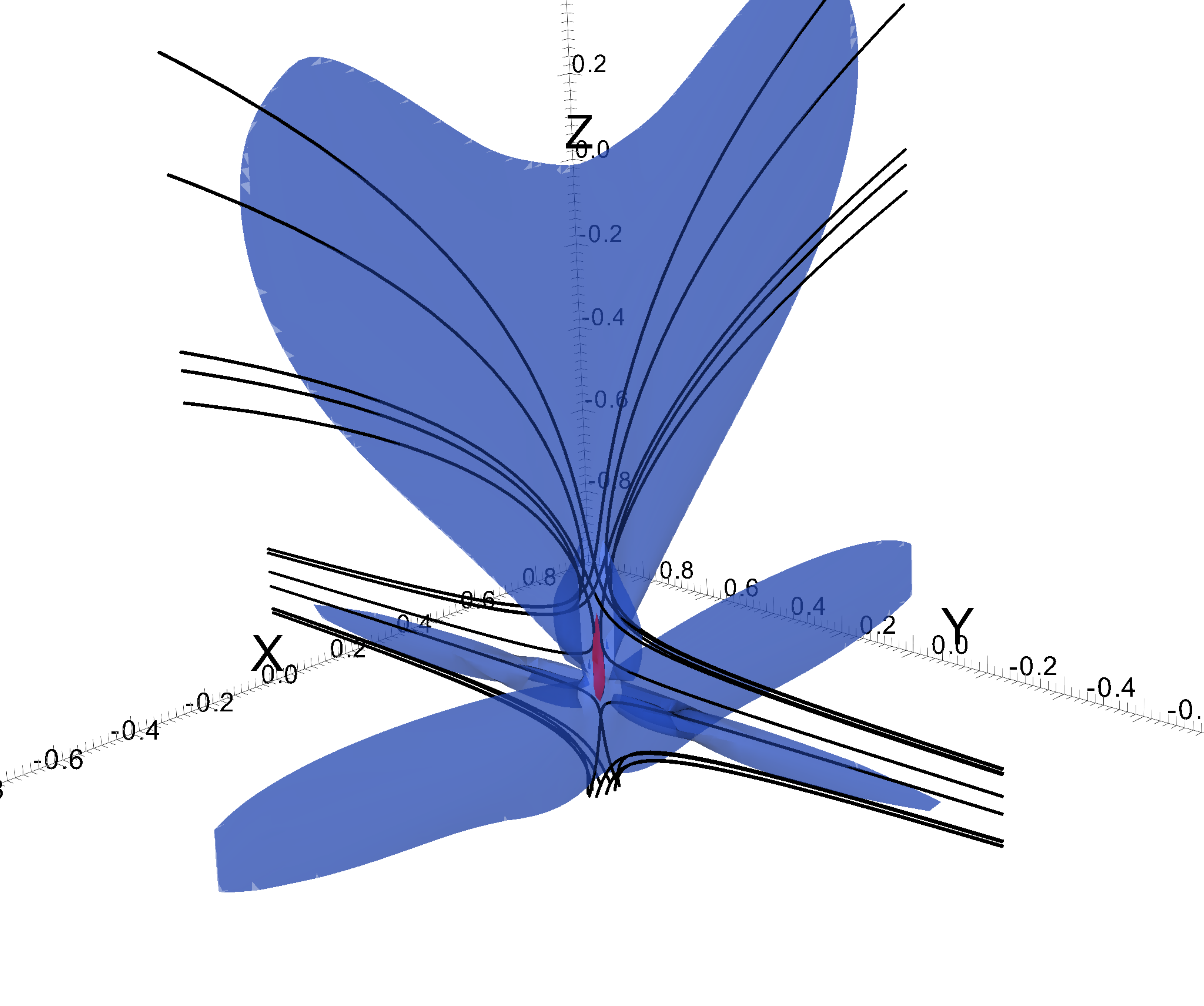}
\caption{Current density distribution and field lines in the relaxed
  state for $b=0.3$ and $N=81$. The blue contour gives an
  isosurface of $J=0.1$ while the red contour is for $J=1$. The field
  lines show the QSL structure and the topology around the null at
  $z=-0.75$ where the strongest currents accumulate.}
\label{fig:currents_b03}
\end{figure}
In the runs of Figure~\ref{fig:psedotime} the null point lies inside
the computational domain ${\cal D} $ at the point $(0, 0, -3/4)$,
following from our parameter choice $b = 0.3$, $\mu = -0.4 $.  We expect
therefore to see current density distributions concentrated about the
null. Figure~\ref{fig:currents_b03} confirms this expectation, as
illustrated by the red isosurface. The black field lines give the fan
structure around the null, aligned roughly to the $y$-$z$-plane. The
blue isosurface legs of weaker current, aligned to the $x$- and $y$-axis,
respectively, indicate that spine and fan currents form simultaneously
(cf. the exposition in \S\ref{sec:spine-fan-qsl}). Further away from
the null, in the upper part of the domain, the field lines illustrate
the additional QSL structure of the field, associated with steep field
line gradients aligned to the $z$-axis. The blue isosurface confirms
the simultaneous formation of currents along the QSL with a magnitude
similar to the fan and spine currents.

\begin{figure}
\includegraphics[width=0.49\textwidth]{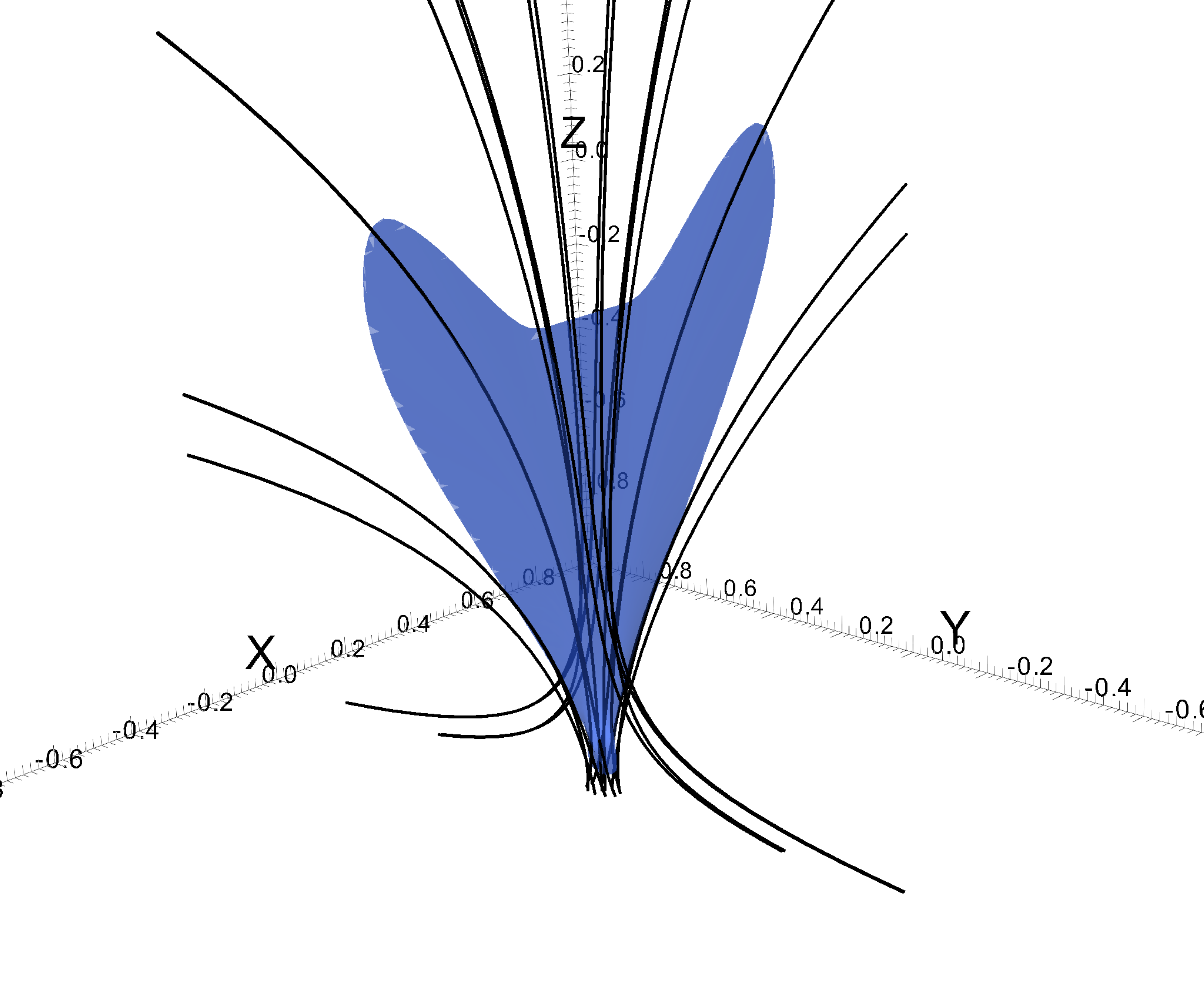}
\caption{Similar to Figure~\ref{fig:currents_b03} but now with
  $b=0.5$, i.e. no null present in the domain. The QSL structure is
  still visible, but the current is weaker and more broadly
  distributed only along the QSL.}
\label{fig:currents_b05}
\end{figure}
In Figure~\ref{fig:currents_b05} we have changed the axial field
parameter to $b = 0.5$ so that the null now lies outside of the
computational box. The QSL structure is still well represented and a
salient feature is the ``exponential'' squashing the field lines
towards the base, as suggested by the $C_4$ field lines of
Eq.~(\ref{eq:fieldeqnull}). The currents are now considerably weaker
than the peak current at the null of the previous setup. Since there
is no preferred location along the QSL, the currents are more evenly
distributed, which can be expected, however, not only from the absence
of a null, but also from the increased strength of the line-tied axial
field which tends to resist current localisation.

\subsection{Scaling of the current density} 
\label{sec:scaling}
\begin{figure}
\noindent\includegraphics[width=0.49\textwidth]{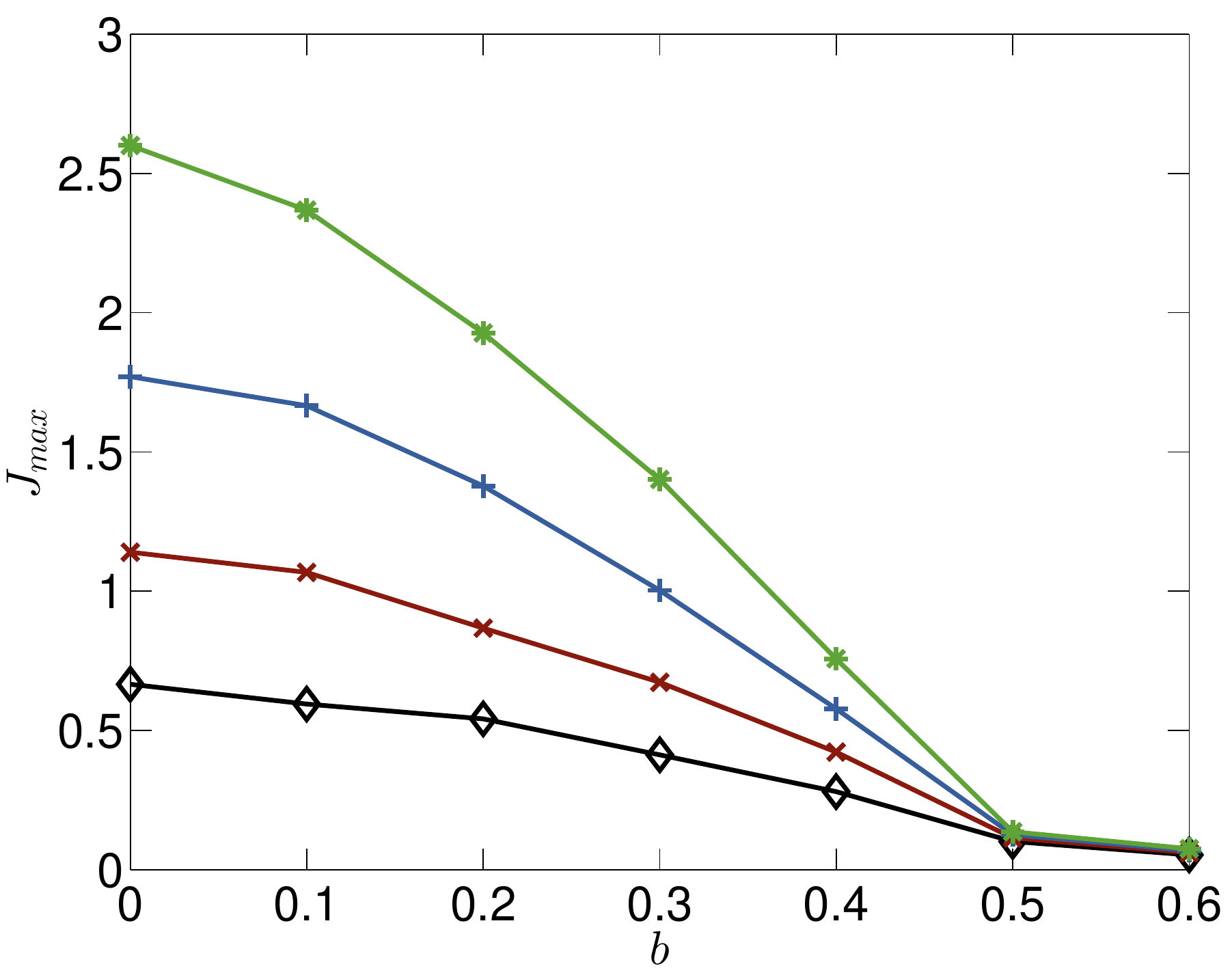}
\caption{Current density $J_{max}$ against $b$ for four different
  resolutions: $N=31$ (black, diamond), $N=41$ (red, x), $N=51$ (blue, +),
  $N=61$ (green, *).}
\label{fig:scaling_b}
\end{figure}
To assess more systematically the influence of the axial field
parameter $b$ on the maximum relaxed current density, we performed a
series of runs with varying $b$ and resolution $N$ (keeping $\mu =
-0.4$ in all cases). Figure~\ref{fig:scaling_b} shows the dependence
of $J_{max}$ on the position of the null. The strongest currents
develop for the weakest axial field and simultaneously largest
distance of the null from the line-tied boundary. Once the null is
outside the domain ($b=0.5$) there is a visible change in the
qualitative behavior of the current formation, which is a direct
result of the field structure as already discussed in the previous
section. The resolution dependence of the current is much weaker for
these cases as well.

\begin{figure}
\noindent\includegraphics[width=0.49\textwidth]{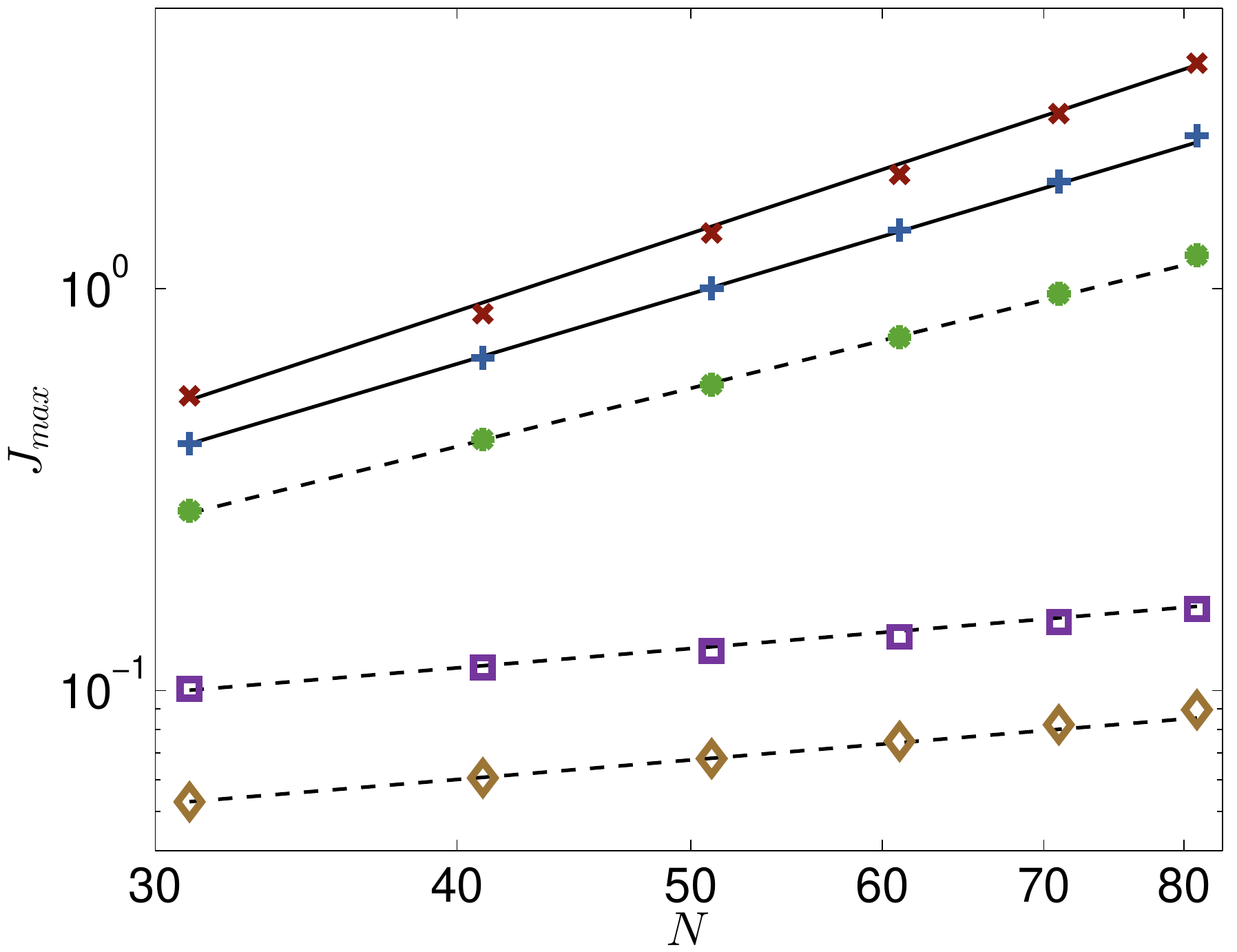}
\caption{Current density $J_{max}$ against $N$ for $b=0.2$ (red, x),
  $b=0.3$ (blue, +), $b=0.4$ (green, *), $b=0.5$ (purple, box),
  $b=0.6$ (brown, diamond). The solid lines indicate fits with strong
  scalings of $J_{max}\propto N^{2.0}$ and $J_{max}\propto N^{1.8}$,
  when the null is in the domain ($b=0.2$ and $b=0.3$), and $J_{max}\propto
  N^{1.5}$ where the null is right at the boundary ($b=0.4$). The
  dashed lines give a weaker scaling fit of $J_{max}\propto N^{0.5}$ for the
  two cases where the null is outside of the domain ($b=0.5$ and
  $b=0.6$).}
\label{fig:scaling_N}
\end{figure}
The quantitative scaling with resolution $N$ for different values of
$b$ and thus null point positions is given in
Figure~\ref{fig:scaling_N}. This illustrates the variation of the
relaxed maximum current density for a sequence of axial field
strengths, specifically $b = 0.2$ to $0.6$ in increments of
$0.1$. Since $\mu = -0.4$ the null is buried below the lower surface
of ${\cal D}$ for the two runs where $ b \ge 0.5 $. We see a strong
scaling with resolution close to $J_{max} \propto N^2 $ for the cases
where the null is actually in the domain, with $b=0.4$ being the
marginal case. For the QSL only current formation, we only find a weak
but still significant scaling of $J_{max} \propto N^{1/2}$.

\subsection{Relation to reconnection models}
The question of what scaling should be expected for the peak current
density with resolution is a key issue, given that ``fast''
reconnection, i.e. reconnection independent of the weak coronal plasma
resistivity, is thought to be required in solar flares. In fact the
$J_{max}$ versus $N$ scalings of the previous section cannot, by
themselves, provide reconnection rates. They can, however, be
interpreted in the light of fast reconnection models. In this case,
the strong $J_{max} \propto N^{2}$ scaling can be shown to be
consistent with reconnection models in which the current sheet
thickness scales linearly with the plasma resistivity
\citep{Petschek-1964,McClymont-Craig-1996}.

This is supported also by analytical reasoning in the simple case of a
collapsing one-dimensional current sheet (with no axial field) modeled
using a Lagrangian description \citep{Craig-Litvinenko-2005}. We find
in our numerical experiments that even for the fully three-dimensional
fields under consideration here, the limiting scalings $J_{max}
\propto N^{2}$ are reasonably approximated for cases where the null is
well situated within the domain and are thus compatible with fast
reconnection models. Conversely, it is hard to reconcile weaker
scalings $J_{max} \propto N^{0.5}$ with any known models of fast
reconnection. It will be interesting to see, if the weak scaling
results persist for e.g. QSL fields constructed from submerged
monopoles.

\subsection{The force-free field} 
\begin{figure}
\includegraphics[width=0.49\textwidth]{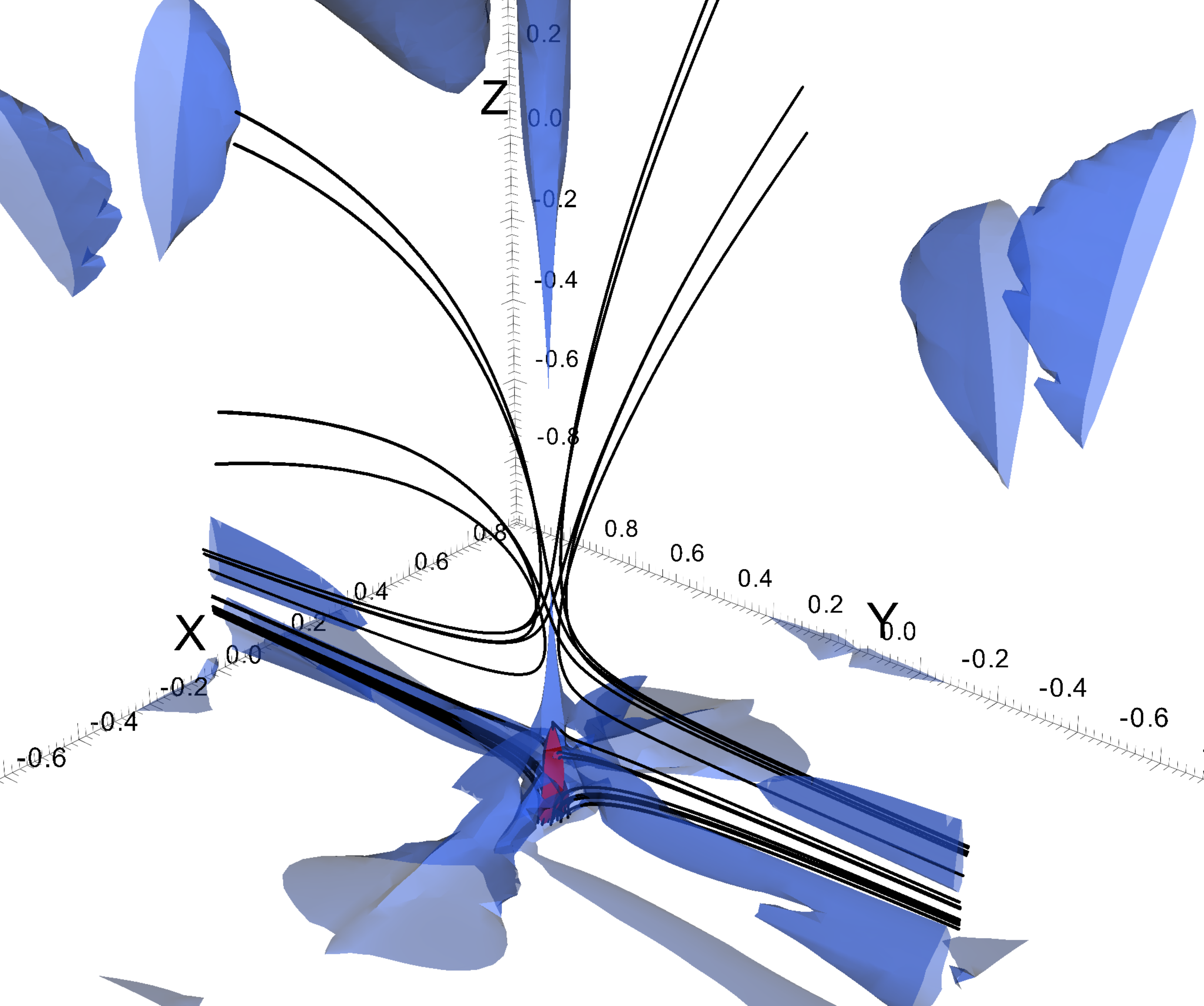}
\caption{Current density distribution and field lines in the relaxed
  state for the initial force-free field (Eq.~\ref{eq:B})
  configuration with $\kappa=-2.5$, $\mu=-0.4$, $b=0.3$, and
  $N=61$. The blue contour gives an isosurface of $J=0.8$ while the
  red contour is for $J=5$. The field lines show again the QSL structure and
  the topology around the null at $z=-0.75$.}
\label{fig:currents_full-field}
\end{figure}
We now consider the full force-free field configuration given by
Equation~\ref{eq:B}. Figure~\ref{fig:currents_full-field} gives the
resulting relaxed current configuration and magnetic field structure
for a computational run with $\kappa=-2.5$, $\mu=-0.4$, $b=0.3$, and
$N=61$. These values were chosen to give a close as possible
configuration to the linear field studies presented above. As can be
seen in the figure, the current again accumulates at the null, which
is present at the same position as before, i.e.  $(0, 0, -3/4)$. The
field structure around the null is also very similar, as should be
expected from the linearization. The QSL currents, however, seem to be
suppressed and overshadowed by the initial currents of the unperturbed
field.  Despite these differences, the field structure still shows the
strong QSL field-line connectivity gradient.  Exploratory computations
with different resolutions indicate that the current magnitude at the
null grows, as previously found in the linear field case, and
reproduces a scaling $J_{max}\propto N^2$ which is again consistent with
fast reconnection. The analysis of current structure and formation is,
however, complicated due to the currents already present in the
initial force-free configuration. We can see nonetheless that our
linearized potential field model derived from the more complex
force-free field can describe the structures and current build-up
close to the null with high accuracy.

\section {Conclusions}
\label{sec:discussion}
We have investigated current formation in bounded line-tied magnetic
field configurations using an ideal magneto-frictional relaxation
method. Our initial fields include potential fields that derive from a
general force-free field configuration. This approach allows us to
distinguish {\it a priori} between fields that contain a magnetic null
in the computational domain ${\cal D}$ and those which comprise only a
strong QSL structure.

For parameter sets that comprise a magnetic null, we find that,
although the strongest currents are always attracted to the null,
significant current layers can still form along the superimposed QSL
structure. The QSL currents, however, are more strongly pronounced in
the simpler, linear potential field than in the considerably more
complicated structures of our general force-free field. The current
formation along the QSLs becomes increasingly prominent for runs in
which the null is positioned below the lower surface of the
computational domain.  For our range of parameters, the currents
appear to spread evenly over the QSL.  We believe this absence of
focusing is due to a lack of an additional structure in the QSL at
least in the fields we examine.  This contrasts to e.g. the hyperbolic
flux tubes in some of the submerged monopole models
\citep{Aulanier-etal-2006,Effenberger-etal-2011}.  It would be of some
interest therefore to extend our investigation to similar monopole
field configurations.  This requires a careful study of the initial
conditions in the relaxation since the strong gradients at the
boundary overlying the monopoles can prevent reconnective currents
from localizing convincingly in the interior of the domain.

One advantage of the present relaxation scheme is that we can follow
current formation in a strictly ``ideal'' (i.e. resistivity-free)
fashion. This allows one to compute scaling laws for the current
divergence against resolution. In particular, we have seen that when
the null is centred in the computational domain, scalings can be
derived $J_{max} \propto N^2$ that are consistent with fast reconnection.
The downside of this approach is that, having no access to the actual
dynamic evolution towards the relaxed state.  we can not draw any
definite conclusions on the time-dependence of the current build up.
Thus, dynamic effects like alignment between the velocity and magnetic
field \citep{Grauer-Marliani-2000} that may come into play in the full
MHD problem are not represented.  We intend to investigate these and
other dynamic effects further in the future, by comparing results from
the relaxation method as employed in this study with three-dimensional
ideal MHD calculations.

\acknowledgments{\small We thank the referee for useful comments which
  helped to improve the manuscript. This work was partially supported
  by the Marsden Fund of New Zealand.}

\end{document}